\documentclass[conference]{IEEEtran}
\ifCLASSINFOpdf
   \usepackage[pdftex]{graphicx}
   \DeclareGraphicsExtensions{.pdf,.jpeg,.png}
\else
\fi
\usepackage{makecell}
\usepackage[utf8]{inputenc}

\begin{document}
%
\title{Towards Reduced Reference Parametric Models for Estimating Audiovisual Quality in Multimedia Services}

\author{\IEEEauthorblockN{Edip Demirbilek}
\IEEEauthorblockA{Institut National de la Recherche Scientifique\\
Montréal, QC, CANADA H5A 1K6\\
E-mail: edip.demirbilek@emt.inrs.ca}
\and
\IEEEauthorblockN{Jean-Charles Grégoire}
\IEEEauthorblockA{Institut National de la Recherche Scientifique\\
Montréal, QC, CANADA H5A 1K6\\
E-mail: gregoire@emt.inrs.ca}}


%


\maketitle

\begin{abstract}
We have developed reduced reference parametric models for estimating perceived quality in audiovisual multimedia services. We have created 144 unique configurations for audiovisual content including various application and network parameters such as bitrates and distortions in terms of bandwidth, packet loss rate and jitter. To generate the data needed for model training and validation we have tasked 24 subjects, in a controlled environment, to rate the overall audiovisual quality on the absolute category rating (ACR) 5-level quality scale. We have developed models using Random Forest and Neural Network based machine learning methods in order to estimate Mean Opinion Scores (MOS) values. We have used information retrieved from the packet headers and side information provided as network parameters for model training. Random Forest based models have performed better in terms of Root Mean Square Error (RMSE) and Pearson correlation coefficient. The side information proved to be very effective in developing the model. We have found that, while the model performance might be improved by replacing the side information with more accurate bit stream level measurements, they are performing well in estimating perceived quality in audiovisual multimedia services.
\end{abstract}

%
\IEEEpeerreviewmaketitle

\section{Introduction}
One way of improving multimedia streaming or real-time multimedia communications is tuning the control parameters to improve Quality of Service (QoS) factors such as maximizing system throughput, reducing response time, decreasing packet loss ratio and jitter value. This kind of approach implicitly aims to improve the overall perceived service quality. However in recent years there has been a shift towards maximizing the perceived quality itself by tuning the media and channel control parameters.

The standard way of measuring perceived quality is by conducting subjective tests where the end user is asked to rate the system as a whole or rate individual system components on a continuous or discrete scale. However defining service quality by conducting subjective tests is resource intensive and not feasible in real-time communications. This raises the question if we can measure or estimate perceived quality automatically with acceptable resource requirements and high accuracy. To answer this question, we need a reference point, in the form of databases of audiovisual material at several quality levels to enable us to develop models for accurate quality estimation. However audiovisual testing is a topic still relatively under-explored and the number of existing databases of various configurations is very limited. In this research, we present a database of audiovisual material with subjective scores for a wide range of audiovisual quality parameters as one of our contributions.

We have taken a parametric model approach by using the additional side information available that makes it possible to develop a model meeting the real-time requirements.  We attempt to estimate the audiovisual quality directly from the influence factors by creating the model with machine learning algorithms that have been successfully applied to estimating the perceived quality. We have mainly used the Random Forests ensemble methods during model training. However we have provided the comparative results with Multi-layer Perceptron (MLP) methods that have been widely used in image assessment, video assessment, and video and voice quality estimation.

The rest of this paper is organized as follows. In section 2, we present standards frequently used while conducting a research similar to ours. In the same section, we briefly explain various perceived quality modelling approaches, statistical metrics used for evaluating the model performances, and best practices used in machine learning when measuring the test accuracy over a relative small set of data. We list existing audiovisual quality databases and the models generated in the section 3. In section 4 we explain the audiovisual database we have generated in details. In section 5 we introduce parametric models that we have developed and share the results obtained for Random Forests and Neural network methods. We sum up the work and share the future work in the conclusion section.

\section{Background}

In this section we briefly discuss the standard practices for audiovisual quality tests, different approaches for developing audiovisual quality models, statistical metrics used for evaluating the model performances, and some machine learning best practices often used when validating models over a small data set.

The International Telecommunication Union (ITU) has various recommendations on how audio, speech, video, and audiovisual quality tests should be conducted. Commonly used recommendations for audiovisual quality tests are P.911 \cite{recommendation1998911} and P.920 \cite{itu1996920}. These recommendations describe the different test methods, provide guidance on the test material to be used, describe test environment, and specify number of subjects and possible subject screening \cite{raake2011ip}.

As per methodology, single-stimulus methods such as the Absolute Category Rating (ACR) are used commonly for collecting subjective quality judgements since they reflect the everyday usage situations of multimedia services better compared to other available methods such as Degradation Category Rating (DCR), BS.1116, Multiple Stimuli with Hidden Reference and Anchor (MUSHRA), or Subjective Assessment Methodology for Video Quality (SAMVIQ). The ACR method enables efficient realistic rating of several files in a session with repeatability. Two widely used ACR scales are 11-point and 5-point scales. Categorical 5-point scale is widely used in telecommunication field where the labels ``excellent'', ``good'', ``fair'', ``poor'',  and ``bad'' translated to the values 5, 4, 3, 2, and 1 when calculating the MOS \cite{garcia2014parametric}.

Audiovisual quality depends on the audio quality, the video quality, their interaction, and audiovisual impairments. The simplest solution for estimating audiovisual quality is to use a function of audio and video quality estimates and compute the audiovisual quality score regardless of what type of degradation affects the audio and video scores. More detailed predictions can be achieved by using intermediate audio and video features that underlie the overall audio and video scores, and map these to an audiovisual quality score. The increased accuracy, however, comes with an increased complexity of the model \cite{raake2011ip}.

There are different approaches in developing the audiovisual quality models. Raake et al \cite{raake2011ip} categorized the models in terms of the level at which input information is extracted.  These quality models exploit the packet header-, bit stream-, or signal-information for providing audio, video, and audiovisual quality estimates.

Signal-based and bit stream models rely on partially or fully decoded received payload. These methods perform well in terms of accuracy but require high processing demands and fall short when the media stream is encrypted \cite{maki2013reduced}.

Parametric audio, video, and audiovisual quality models are developed and standardized in ITU-T Study Group 12 under the provisional name P.NAMS. Garcia and Raake \cite{garcia2013parametric} \cite{garcia2014parametric},  demonstrated that parametric models can be tuned to specific use case very accurately with thorough understanding of the content delivery protocols and selecting the appropriate side information \cite{maki2013reduced}.

Additionally, the quality models can be classified according to the amount of information they need from the original signal. In no reference (NR) models, no information from the original signal is used while full reference (FR) models have access to the original source sequence, which is compared with the processed sequence. Reduced reference (RR) models use the processed sequence together with a set of parameters extracted from the source sequence.

The performance of a quality model is evaluated using statistical metrics. Traditionally, three statistical metrics are used to report the model performance's accuracy, consistency and linearity/monotonicity \cite{garcia2014parametric}.

An accurate model aims to predict the subjective quality scores with the lowest error in terms of Root Mean Square Error (RMSE), which depends on the rating scale, used during the subjective tests.  ITU-T Recommendation P.863 \cite{beerends2013perceptual} recommends to convert this value to the so-called epsilon-modified Root-Mean-Square-Error (RMSE\*) to compare the results across different scales \cite{garcia2014parametric}.

The perceived quality predictions have to have consistently low error margin over the range of test subjects and agree with the relative magnitude of subjective quality ratings. The model's consistency is reported by computing either the outlier ratio or the residual error distribution. The outliers are defined as the points for which the prediction error exceeds the 95\% confidence interval \cite{garcia2014parametric}.

In the literature there are two commonly used metrics for computing the linearity of a model; Pearson Correlation coefficient and Spearman Rank coefficient. The Pearson correlation coefficient is used when the data is drawn from a test set with near-normal distribution. In other cases the Spearman Rank coefficient is used to report the linearity between the estimated and the actual subjective quality scores \cite{garcia2014parametric}.

The quality scores estimated by the quality model are compared to the quality scores obtained from subjective tests in order to assess the performance of the quality models. However, in the case of limited amount of training and test data set, K-fold or leave-one-out cross-validation are used to report the performance of the quality model. In the K-Folds approach, available data is split into K folds. In each step, K-1 folds are used to train the data and the remaining 1 fold is used to measure the accuracy of the model. This procedure is repeated K times by using a different portion of the available data as test data. Data splitting can be done randomly as well as stratifying the folds. In order to make the predictions more robust and independent of the selected K folds, the whole procedure is typically run several times and the average of these runs is taken for each metric. The common practice is to use stratified 10-fold cross-validation \cite{garcia2014parametric}.

\section{Related Work}

The QUALINET Multimedia Databases set v5.5 \cite{fliegel2014qualinet} provides a list of some publicly available audiovisual databases and models.

In 2010 Goudarzi et al \cite{goudarzi2010audiovisual} conducted subjective experiments to explore methods to predict audiovisual quality objectively for video calls in wireless applications. They have presented subjective test results for 60 test conditions on how audio and video contribute to overall audiovisual quality and develop models to reflect this relationship, and investigated how network and application parameters  affect overall audiovisual quality. In their analysis they have used a regression model to predict audiovisual quality from packet loss rate and frame rate.


The Video Quality Experts Group (VQEG) ran subjects tests through the same audiovisual material in six different international laboratories \cite{pinson2012influence} \cite{pinson2013subjective}. Each of these six labs conducted the experiment in a controlled environment while four labs also repeated the same experiments in a public environment. They have reported that audiovisual subjective tests are highly repeatable from one laboratory and environment to the next and recommended 24 or more subjects for ACR tests in a lab and 35 and more subjects for the same sensitivity in the public environment.

Robitza et al \cite{robitza2012made} have presented a video database especially designed for mobile TV quality assessment. They have reported the results of a broad study to provide content creation and editing guidelines adapted for mobile TV and we have analyzed the impact of these guidelines on the perception of quality through a subjective experiment.

Mäki et al \cite{maki2013reduced} developed a reduced reference parametric model for audiovisual quality estimation following the Pseudo-Subjective Quality Assessment (PSQA) methodology and have compared different kinds of statistical estimators, namely Multilayer Perceptrons (MLP) and Random Neural Networks (RNN). They have trained the model with subjective assessment data for an IPTV-like scenario and have reported that the model generated performs at its best when implemented with a MLP. They have also showed that that by adding a small amount of information about the original signal, the performance of these packet-level models can be very good for certain uses.

In \cite{garcia2013parametric} \cite{garcia2014parametric}, the authors presented the winning model of the P.NAMS (“Parametric Non-intrusive Assessment of audiovisual Media Streaming quality”) competition for the Higher Resolution application area. Their model focuses on a non-intrusive parametric packet-based audiovisual quality model and standardized as the ITU-T Recommendation P.1201.2. Their model was based on the results of 10 and validated on 14 subjective tests that cover typical audio and video degradations for IPTV, including audio and video compression artifacts and packet loss, slicing and freezing artifacts. They have obtained high performance results with Pearson Correlation r=0.911 and RMSE =0.435 for the audiovisual model, with r =0.902 and RMSE =0.461 for video, and with r =0.949 and RMSE =0.336 for audio on the 5-point scale used.

All of these audiovisual databases have a variety of configurations. Researchers have developed audiovisual quality models using the data provided in these databases. Some models aim to estimate the perceived quality directly  by conducting the audio-video subjective tests while others conduct audio, video and audio-video test separately and try to deduct a model accurate enough for estimating the audiovisual quality by using the separate audio and video quality estimates as parameters.

As recent developments show, reduced reference parametric models achieve high accuracy in estimating the perceived quality with limited resources. In this work we try to build a similar model by estimating the audiovisual quality directly by using Random Forest and Neural network machine learning methods for specific target network configurations.

\section{An Audiovisual Quality Database}
Creating an audiovisual quality database require finding optimum database configurations, developing required test tools, building test setup, producing the content,  generating the files under specific network conditions, preparing the subjective test methodology and conducting the tests while following the standards. Publicly available databases help us to avoid this  significant amount work and enable us to compare the performance of various models on the same data set. However eventually a model performance has to be measured on the target environment with the required application and network parameters. Additionally to our knowledge none of the available databases include the whole parameter space we needed to experiment.

\begin{figure}[ht]
\centering
\includegraphics[width=3.2in]{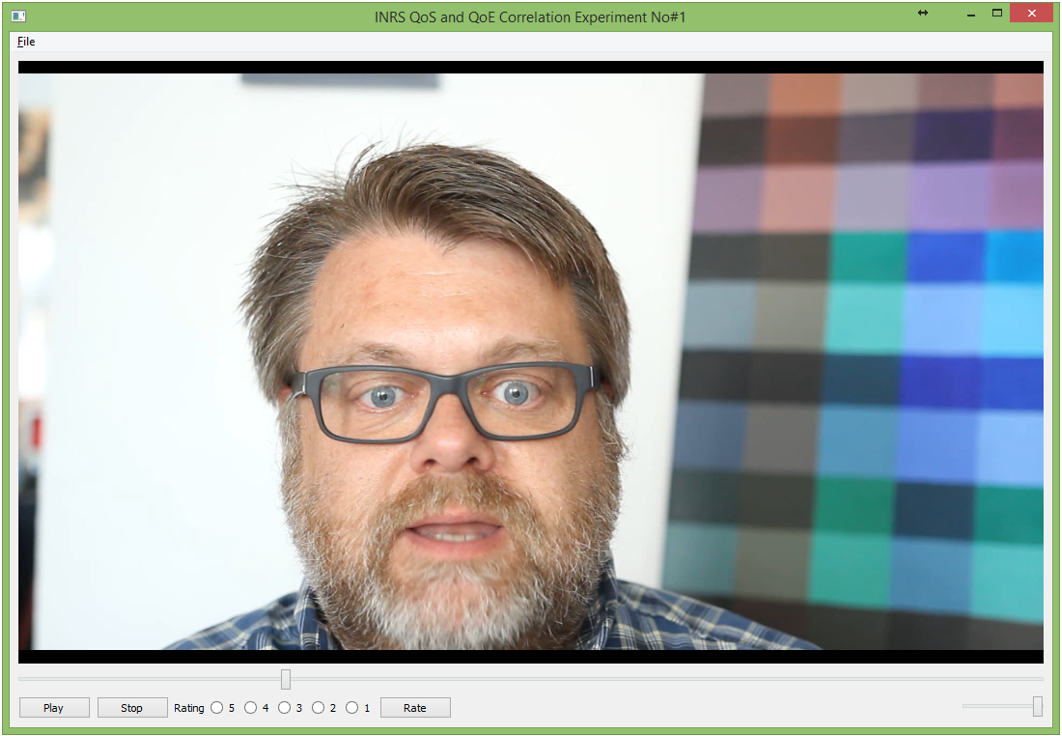}
\vspace{-0.2cm}
\caption{A scene from the generated reference video file.}
\label{fig:scene}
\vspace{-0.2cm}
\end{figure}

We have created an audiovisual content specifically for this research. Figure \ref{fig:scene} depicts a scene in the generated reference video being played by the custom video player developed to collect the subjective scores without disclosing the resource’s quality. The video consists of slow-moving scenes where a person reads a passage from a book. This content is chosen to be similar to a typical one-to-one audiovisual conversation.

We have created an audiovisual database in order to develop and validate a reduced reference parametric model to estimate the user perceived audiovisual quality in multimedia services. The database is designed to include the resolution, bit rate, bandwidth, packet loss rate, and jitter influence factors. Table \ref{tab:influence_factors} shows the selected values for these influence factors. Each of these influence factor values is assigned as follows.

\begin{table}[ht]
\caption{Audivisual quality database influence factors}
\label{tab:influence_factors}
\centering
\begin{tabular}{|c||c|}
\hline
\thead{Resolution} & \makecell{HD1080 (1920x1080 pixels), \\ HD720 (1280x720 pixels)} \\
\hline
\thead{Bit Rate} & \makecell{High Quality(HQ), \\ Middle Quality(MQ), \\ Low Quality (LQ)} \\
\hline
\thead{Bandwidth} & \makecell{High Bandwidth (2x Max Bit Rate), \\ Low Bandwidth (Max Bit rate)} \\
\hline
\thead{Packet Loss rate (\%)} & 0, 0.1, 0.5 \\
\hline
\thead{Jitter (ms)} & 0, 10, 50, 100 \\
\hline
\end{tabular}
\vspace{-0.2cm}
\end{table}

From the reference video file, we have created 6 MPEG audio-video files with a leading commercial video editing software. These 6 source files consisted of 3 target bit rates (High Quality, Middle Quality and Low Quality) for two resolution levels. Considering the adaptation to the mobile platforms HD 1080 and HD720 resolution levels are selected. Table \ref{tab:source_files_generated} shows the source files generated where the bitrates computed as an average over one second of transmission. Each source files are encoded with MPEG video version 2 video codec with Main@High 1440 profile and MPEG Audio version 1 layer 2 audio codec contained in an MPEG-TS container. Video frame rate are set to 25 fps and audio sampling rate set to 48.0 KHz.

\begin{table}[ht]
\caption{Source Files Generated.}
\label{tab:source_files_generated}
\centering
\begin{tabular}{|c||c||c||c|}
\hline
\thead{File} & \thead{Overall \\ BitRate \\ (Kbps)} & \thead{Video \\ Max \\ BitRate \\ (Kbps)} & \thead{Audio \\ BitRate \\ (Kbps)} \\
\hline
MPEG2\_HD\_720\_LQ.ts & 1389 & 1477 & 128 \\
\hline
MPEG2\_HD\_720\_MQ.ts & 3461 & 3664 & 128 \\
\hline
MPEG2\_HD\_720\_HQ.ts & 8040 & 8313 & 128 \\
\hline
MPEG2\_HD\_1080\_LQ.ts & 2871 & 3227 & 128 \\
\hline
MPEG2\_HD\_1080\_MQ.ts & 7457 & 8069 & 128 \\
\hline
MPEG2\_HD\_1080\_HQ.ts & \makecell{13.1 Mbps} & 18083 & 128 \\
\hline
\end{tabular}
\vspace{-0.2cm}
\end{table}

In our preliminary experiments we have selected packet loss rate (PLR) between 0-5\% and observed that the PLR greater than 0.5\% reduces the perceived quality significantly which is never intended in real-life scenarios. We have obtained similar results with the bandwidth configurations. Initially we have tested 4 different bandwidth levels for a given bit rate configuration and observed that only two out of these 4 levels are relevant to the real-life conditions and eventually only kept these two for our study.

These selected bandwidth-pair configurations are intended to provide one configuration for no-limitation on bandwidth (High Bandwidth) while the other one is intended to provide slightly less bandwidth than max bitrate (Low Bandwidth), which causes only small degradation in perceived quality.

We have conducted various iperf tests to measure the effective bandwidth in Low Bandwidth cases and find out that available bandwidth being 2.8\% less than max bitrate for each file. We have taken into account that iperf adds a small bias to the measurement since iperf measures the available bandwidth using a TCP stream, while the bandwidth limitation sets the bandwidth available for IP packets \cite{nussbaum2009comparative}.

The test sequences were prepared prior to the assessment by recording RTP-based video streams transmitted over an emulated network. The videos were streamed and recorded with the VideoLan VOD Server and VLC media player. The netem network emulator is deployed in order to introduce the packet loss and jitter test conditions. Dummynet is used to manage the bandwidth settings between the VOD server and the client. A total of 144 network conditions were considered and respectively 144 audio-video files are recorded for subjective quality tests.

The participants consisted of 24 graduate level INRS students that are familiar with the multimedia quality assessment, subjects coming from various backgrounds with a fluency in English language selected for delivering the test guidelines and for answering any questions raised during the training session. The subjects' age ranged between 20 and 37 years old.

The viewing and listening conditions specified in P.911 \cite{recommendation1998911} were followed as much as feasible. Subjects were asked to rate each audio-video quality on the 5-point ACR categorical quality scale.  Subjects were allowed to submit their subjective scores after watching/listening the first 10 seconds. The order of the rendered sequences was randomly drawn before assessment but was the same for all subjects. Instructions are given in English and subjects were allowed to ask questions. Subjects initially performed a training session and completed the tests between 30-45 min in a single assessment session. Subjects were allowed to have a pause halfway through the test.

\section{Two reduced reference parametric models}

The models we mention in this section are trained on the 5-point ACR MOS scale where the scores for a given audiovisual configuration are averaged over all subjects. We have constructed various RR parametric models and measured their performance in terms of accuracy, consistency and linearity. We have seen earlier that these terms are represented by the following statistical metrics; Root-Mean-Square-Error (RMSE), the outlier ratio which is typically defined as the points for which the prediction error exceeds the 95\% confidence interval, and the Pearson correlation coefficient.

We have extracted the features from the file headers such as bits per pixel in each video frame, audio video delay, duration, frame count, video and audio stream sizes… and additional side information such as network packet loss, network jitter and bandwidth configurations. We have kept the feature space the same across all machine learning models we have tried.

In our quick Weka \cite{witten2005data} experiments, we have witnessed an overall superior performance of the decision tree based ensemble methods. Out of all available ensemble methods, Random Forests showed better accuracy in terms of RMSE values calculated. Neural networks have been widely used in audio and video quality estimation. In order to put the Random Forest model's performance into relation to neural networks based models, we have decided to develop two models based on Random Forest and Neural Networks. Mäki et al \cite{maki2013reduced} showed that Multi-layer perceptron (MLP) models perform better compared to Random Neural Network (RNN) models. Therefore as neural network implementation, we have used Multi-layer Perceptron.

First we have trained an MLP regression model using a single hidden layer where the number of input neurons equalled the number of input features. The tangent-hyperbolic function was chosen as the activation function of the hidden nodes, while linear function was chosen for the output neuron as in \cite{maki2013reduced}. The learning rate was set to 0.02 and the number of iterations was set to 100 for gradient descent to perform on the neural network's weights.

Second we have trained a Random Forests regression model that fits a number of classifying decision trees on various sub-samples of the dataset and use averaging to improve the predictive accuracy and control over-fitting. The number of trees in the forest was set to 100 with no restriction on the depth of the tree and all features are used.

Initially the dataset set was shuffled and then both methods were trained and their accuracy is measured on the test MOS data using 10-Fold cross validation. To reduce the variation, as a common practice, we have run this process 10 times and have taken the average of the measured statistical metrics. These figures are shown in Table \ref{tab:performance}. It is clear that Random Forests methods outperform Multi-layer perceptron methods in terms of all metrics computed.

\begin{table}[ht]
\caption{Random Forests vs Multi-layer Perceptron Performance}
\label{tab:performance}
\centering
\begin{tabular}{|c||c||c||c|}
\hline
\thead{Algorithm} & \thead{RMSE} & \thead{Pearson \\ Correlation} & \thead{95\% Confidence \\ Interval} \\
\hline
\makecell{Random Forests} & 0.3138 & 0.8871 & 0.597 \\
\hline
\makecell{Multi-layer Perceptron} & 0.4207 & 0.8023 & 0.767\\
\hline
\end{tabular}
\vspace{-0.2cm}
\end{table}

\begin{figure}[ht]
\centering
\includegraphics[width=3.2in]{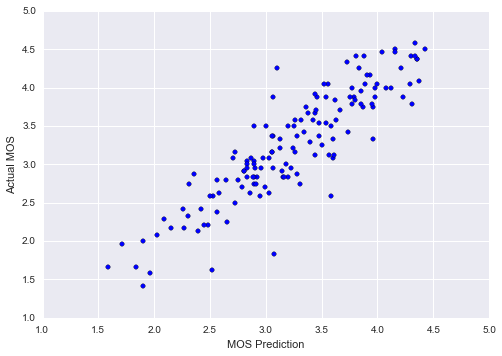}
\vspace{-0.2cm}
\caption{Actual MOS vs predicted MOS: Random Forests}
\label{fig:rf_matrix}
\vspace{-0.2cm}
\end{figure}

\begin{figure}[ht]
\centering
\includegraphics[width=3.2in]{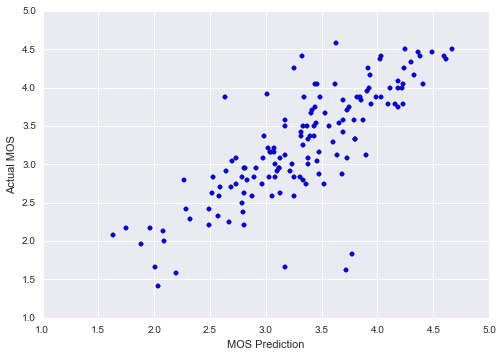}
\vspace{-0.2cm}
\caption{Actual MOS vs predicted MOS: Multi-layer Perceptron}
\label{fig:mlp_matrix}
\vspace{-0.2cm}
\end{figure}

The difference in performances is much easier to realize in graphical interpretation.  In Figure \ref{fig:rf_matrix} and Figure \ref{fig:mlp_matrix} actual MOS vs predicted MOS for both Random Forests and Multilayer Perceptron methods. The figures clearly show that Random Forest method makes visibly more accurate estimation.

\begin{figure*}[!t]
\centering
\includegraphics[width=0.8\textwidth]{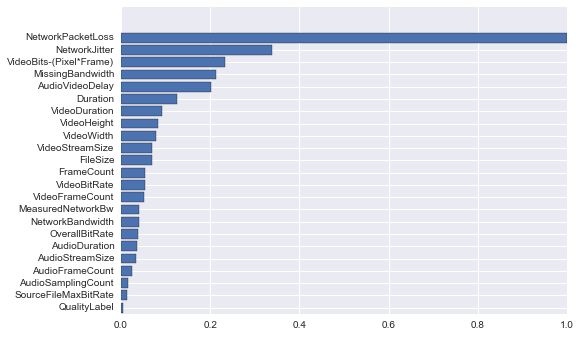}
\vspace{-0.2cm}
\caption{Random Forests Feature Importance: Network PLR, jitter and bandwidth information have the most influence on estimating the perceived quality.}
\label{fig:rf_feature_importance}
\vspace{-0.2cm}
\end{figure*}

We have intentionally kept using all features when training both models. With some feature selection pre-processing the MLP performance might be improved. The beauty of the Random Forests method is that it handles the feature selection automatically as well as tells us the feature importance which would be extremely useful while adapting the service quality based on the quality predictions made.  In Figure \ref{fig:rf_feature_importance} it is shown that packet loss rate, network jitter and bandwidth information, provided as side information, plays the most important role when estimating the perceived quality. It is important to note that changing the range of value of parameters would influence the perceived quality differently. The values here are the side information but not the actual data collected from the bit stream level. This automatically makes it clear that a hybrid approach that incorporates the packet level information with more accurate bit stream level information would produce much better predictions.

The RMSE, Pearson correlation coefficient values reported are close to the values reported by other researchers including the winner of the P.NAMS competition. However when we look at the figures, we clearly see the outlier points where the MOS estimation values differ from actual MOS value by a large margin. After we have carefully analyzed the actual MOS values, we have discovered some of them differ by more than 2 MOS points compared to their actual expected range. Similar issues have been reported by Mäki et al \cite{maki2013reduced} as well. The reason behind this is the difference between the channel parameters provided as side information and the actual bit stream level information. In a hybrid approach, where the packet header information and more accurate bit-stream level information is used, this problem would not occur.

\section{Conclusion and Future Work}
We have developed an audiovisual quality database in order to gather data for developing two reduced reference parametric models for estimating audiovisual quality in multimedia services. We have trained the models using Random Forests and Multi-Layer Perceptron machine learning methods. In terms of RMSE, Pearson Correlation coefficient value and 95\% confidence interval boundaries, Random Forests based methods outperform Multi-Layer Perceptron methods and we have obtained results in the same range as reported by researchers.

Developing a parametric model requires less effort compared to bit stream level or signal level models. Using the side information proved to be invaluable in terms of improving the performance metrics. However the model might suffer from the imperfections contained in the side information.

The reduced reference parametric model based on the Random forests achieved high accuracy. Random Forests methods also provide built-in feature importance properties that give insight about which parameters are more influential on the user perceived service quality. This information would be very useful when adapting the service quality based on the quality predictions made.

Instead of side information which has potential imperfections, bit stream level measurements can be used to obtain more accurate quality estimations. However this approach would require peeking into the bit stream and would require more effort to build such a model.

As part of our ongoing research, next we will concentrate on hybrid modelling approaches where both packet header and bit stream level information is used to develop models and create an actual real-time communication framework where collecting the subjective scores, calculating the MOS values, building the models and reporting the results are all integrated. Eventually we will use the predictions made to maximize the perceived quality itself by tuning the media and channel control parameters.

\bibliographystyle{IEEEtran}
\bibliography{IEEEabrv,IEEEexample}

\begin{thebibliography}{10}
\providecommand{\url}[1]{#1}
\csname url@samestyle\endcsname
\providecommand{\newblock}{\relax}
\providecommand{\bibinfo}[2]{#2}
\providecommand{\BIBentrySTDinterwordspacing}{\spaceskip=0pt\relax}
\providecommand{\BIBentryALTinterwordstretchfactor}{4}
\providecommand{\BIBentryALTinterwordspacing}{\spaceskip=\fontdimen2\font plus
\BIBentryALTinterwordstretchfactor\fontdimen3\font minus
  \fontdimen4\font\relax}
\providecommand{\BIBforeignlanguage}[2]{{%
\expandafter\ifx\csname l@#1\endcsname\relax
\typeout{** WARNING: IEEEtran.bst: No hyphenation pattern has been}%
\typeout{** loaded for the language `#1'. Using the pattern for}%
\typeout{** the default language instead.}%
\else
\language=\csname l@#1\endcsname
\fi
#2}}
\providecommand{\BIBdecl}{\relax}
\BIBdecl

\bibitem{recommendation1998911}
I.~Recommendation, ``911: Subjective audiovisual quality assessment methods for
  multimedia applications,'' \emph{International Telecommunications Union,
  Geneva}, 1998.

\bibitem{itu1996920}
R.~P. ITU-T, ``920, interactive test methods for audiovisual communications,''
  \emph{International Telecommunications Union Radiocommunication Assembly},
  1996.

\bibitem{raake2011ip}
A.~Raake, J.~Gustafsson, S.~Argyropoulos, M.~Garcia, D.~Lindegren, G.~Heikkila,
  M.~Pettersson, P.~List, B.~Feiten \emph{et~al.}, ``Ip-based mobile and fixed
  network audiovisual media services,'' \emph{Signal Processing Magazine,
  IEEE}, vol.~28, no.~6, pp. 68--79, 2011.

\bibitem{garcia2014parametric}
M.-N. Garcia, \emph{Parametric Packet-based Audiovisual Quality Model for IPTV
  Services}.\hskip 1em plus 0.5em minus 0.4em\relax Springer, 2014.

\bibitem{maki2013reduced}
T.~Maki, D.~Kukolj, D.~Dordevic, and M.~Varela, ``A reduced-reference
  parametric model for audiovisual quality of iptv services,'' in \emph{Quality
  of Multimedia Experience (QoMEX), 2013 Fifth International Workshop
  on}.\hskip 1em plus 0.5em minus 0.4em\relax IEEE, 2013, pp. 6--11.

\bibitem{garcia2013parametric}
M.-N. Garcia, P.~List, S.~Argyropoulos, D.~Lindegren, M.~Pettersson, B.~Feiten,
  J.~Gustafsson, and A.~Raake, ``Parametric model for audiovisual quality
  assessment in iptv: Itu-t rec. p. 1201.2,'' in \emph{Multimedia Signal
  Processing (MMSP), 2013 IEEE 15th International Workshop on}.\hskip 1em plus
  0.5em minus 0.4em\relax IEEE, 2013, pp. 482--487.

\bibitem{beerends2013perceptual}
J.~G. Beerends, C.~Schmidmer, J.~Berger, M.~Obermann, R.~Ullmann, J.~Pomy, and
  M.~Keyhl, ``Perceptual objective listening quality assessment (polqa), the
  third generation itu-t standard for end-to-end speech quality measurement
  part i—temporal alignment,'' \emph{Journal of the Audio Engineering
  Society}, vol.~61, no.~6, pp. 366--384, 2013.

\bibitem{fliegel2014qualinet}
K.~Fliegel, ``Qualinet multimedia databases v5. 5,'' 2014.

\bibitem{goudarzi2010audiovisual}
M.~Goudarzi, L.~Sun, and E.~Ifeachor, ``Audiovisual quality estimation for
  video calls in wireless applications,'' in \emph{Global Telecommunications
  Conference (GLOBECOM 2010), 2010 IEEE}.\hskip 1em plus 0.5em minus
  0.4em\relax IEEE, 2010, pp. 1--5.

\bibitem{pinson2012influence}
M.~H. Pinson, L.~Janowski, R.~P{\'e}pion, Q.~Huynh-Thu, C.~Schmidmer,
  P.~Corriveau, A.~Younkin, P.~L. Callet, M.~Barkowsky, and W.~Ingram, ``The
  influence of subjects and environment on audiovisual subjective tests: An
  international study,'' \emph{Selected Topics in Signal Processing, IEEE
  Journal of}, vol.~6, no.~6, pp. 640--651, 2012.

\bibitem{pinson2013subjective}
M.~H. Pinson, C.~Schmidmer, L.~Janowski, R.~Pepion, Q.~Huynh-Thu, P.~Corriveau,
  A.~Younkin, P.~Le~Callet, M.~Barkowsky, and W.~Ingram, ``Subjective and
  objective evaluation of an audiovisual subjective dataset for research and
  development,'' in \emph{2013 Fifth International Workshop on Quality of
  Multimedia Experience (QoMEX)}, 2013.

\bibitem{robitza2012made}
W.~Robitza, Y.~Pitrey, M.~Nezveda, S.~Buchinger, and H.~Hlavacs, ``Made for
  mobile: a video database designed for mobile television,'' in \emph{Sixth
  International Workshop on Video Processing and Quality Metrics for Consumer
  Electronics (VPQM)}, 2012.

\bibitem{nussbaum2009comparative}
L.~Nussbaum and O.~Richard, ``A comparative study of network link emulators,''
  in \emph{Proceedings of the 2009 Spring Simulation Multiconference}.\hskip
  1em plus 0.5em minus 0.4em\relax Society for Computer Simulation
  International, 2009, p.~85.

\bibitem{witten2005data}
I.~H. Witten and E.~Frank, \emph{Data Mining: Practical machine learning tools
  and techniques}.\hskip 1em plus 0.5em minus 0.4em\relax Morgan Kaufmann,
  2005.

\end{thebibliography}

\end{document}